# Discovery of intrinsic ferromagnetism in 2D van der Waals crystals

*Cheng Gong,[1+] Lin Li,[2+] Zhenglu Li,[3,4+] Huiwen Ji,[5] Alex Stern,[2] Yang Xia,[1] Ting Cao,[3,4] Wei Bao,[1] Chenzhe Wang,[1] Yuan Wang,[1,4] Z. Q. Qiu,[3] R. J. Cava,[5] Steven G. Louie,[3,4\*] Jing Xia,[2\*] Xiang Zhang[1,4\*]*

[1]NSF Nano-scale Science and Engineering Center (NSEC), 3112 Etcheverry Hall, University of California, Berkeley, California 94720, USA

[2]Department of Physics and Astronomy, University of California, Irvine, Irvine, California 92697, USA

[3]Department of Physics, University of California, Berkeley, California 94720, USA

[4]Material Sciences Division, Lawrence Berkeley National Laboratory, 1 Cyclotron Road, Berkeley, California 94720, USA

[5]Department of Chemistry, Princeton University, Princeton, New Jersey 08540, USA

[+]These authors contributed equally to this work.

[*]Emails: xiang@berkeley.edu, xia.jing@uci.edu, sglouie@berkeley.edu

**It has been long hoped that the realization of long-range ferromagnetic order in two-dimensional (2D) van der Waals (vdW) crystals, combined with their rich electronic and optical properties, would open up new possibilities for magnetic, magnetoelectric and magneto-optic applications.[1-4] However, in 2D systems, the long-range magnetic order is strongly hampered by thermal fluctuations which may be counteracted by magnetic anisotropy, according to the Mermin-Wagner theorem.[5] Prior efforts via defect and composition engineering,[6-11] and proximity effect only locally or extrinsically introduce magnetic responses. Here we report the first experimental discovery of intrinsic long-range ferromagnetic order in pristine $Cr_2Ge_2Te_6$ atomic layers by scanning magneto-optic Kerr microscopy. In such a 2D vdW soft ferromagnet, for the first time, an unprecedented control of transition temperature of ~ 35% – 57% enhancement is realized via surprisingly small fields (≤ 0.3 Tesla in this work), in stark contrast to the stiffness of the transition temperature to magnetic fields in the three-dimensional regime. We found that the small applied field enables an effective anisotropy far surpassing the tiny magnetocrystalline anisotropy, opening up a sizable spin wave excitation gap. Confirmed by renormalized spin wave theory, we explain the phenomenon and conclude that the unusual field dependence of transition temperature constitutes a hallmark of 2D soft ferromagnetic vdW crystals. Our discovery of 2D soft ferromagnetic $Cr_2Ge_2Te_6$ presents a close-to-ideal 2D Heisenberg ferromagnet for studying fundamental spin behaviors, and opens the door for exploring new applications such as ultra-compact spintronics.**

Atomic layered vdW crystals provide ideal 2D material systems hosting exceptional physical properties. [12-14] Emerging functional devices[14] (e.g., ultrafast photodetector, broadband optical modulator, excitonic semiconductor laser, etc.) have been derived primarily from the electron charge degree of freedom, whereas 2D spintronics[1-3,15] based on vdW crystals is still in its infancy, severely hindered by the lack of long-range ferromagnetic order that is crucial for macroscopic magnetic effects.[4,7] The emergence of ferromagnetism in 2D vdW crystals, if possible, combined with their rich electronics and optics, could open up numerous opportunities for 2D magnetic, magnetoelectric, and magneto-optic applications.[2,3]

The absence of ferromagnetic order in many intensively studied 2D vdW crystals (e.g., graphene or $MoS_2$) has motivated tremendous efforts to extrinsically induce magnetism by defect engineering via vacancies, adatoms, grain boundaries, and edges,[6-11,16-18] by adding magnetic species via intercalation or substitution,[19] or by magnetic proximity effect. In these schemes, extrinsically introduced local magnetic moments are difficult to be correlated in long range via robust exchange interaction,[7] or substrate-induced magnetic responses of 2D materials are rather limited. Theoretical proposals for inducing ferromagnetism by band structure engieering[18,20,21] still await experimental realization. In contrast, if realizable, intrinsic ferromagnetism originating from the parent 2D lattice itself is fundamental for both understanding the underlying physics of electronic and spin processes, and device applications.

Whether or not long-range ferromagnetic order that exists in bulk can persist in 2D regime is a fundamental question, because the strong thermal fluctuations may easily destroy the 2D ferromagnetism, according to the Mermin-Wagner theorem.[5] As illustrated in Fig. 1, down to the root of 2D ferromagnetism, the presence of a spin wave excitation gap – as a direct result from magnetic anisotropy – is a must for long-range ferromagnetic order at finite temperature. Most of the reported ferromagnetic vdW bulk crystals so far are magnetically soft with small easy-axis anisotropy. The challenge of harnessing the long-range ferromagnetic order in 2D vdW crystals hinges on the strength of magnetic anisotropy retained in the 2D regime. In this work, we report the first experimental discovery of long-range ferromagnetic order in pristine $Cr_2Ge_2Te_6$ atomic layers by temperature- and magnetic field-dependent Kerr effect via scanning magneto-optic Kerr microscopy. In our 2D soft ferromagnetic vdW crystal, an unprecedented magnetic field control of ferromagnetic transition temperature of ~ 35% – 57% enhancement is achieved by surprisingly small fields (≤ 0.3 T).

Mechanical exfoliation via scotch tape was applied to prepare $Cr_2Ge_2Te_6$ atomic layers on 260 nm $SiO_2$/Si chips. Bulk $Cr_2Ge_2Te_6$ was reported[22,23] to be ferromagnetic below 61 K, with an out-of-plane magnetic easy axis and negligible coercivity (Fig. S1b,c in supplementary information (SI)). The monolayer (1L) was found to degrade rapidly and become invisible under an optical microscope (Fig. S2), and the peeling off of bilayer (2L) flakes was repeatable and shown to be robust without traceable degradation within 90 minutes in ambient atmosphere (Figs. S4 and S5). Therefore, the thinnest structure we studied in this work is a bilayer. The thickness of the atomic layers is determined by the combination of optical contrast and atomic force microscopy (Fig. S3).

The scanning Kerr microscope used in this work was constructed with a fiber-optic Sagnac interferometer[24] with $10^{-8}$ rad DC Kerr sensitivity and micrometer spatial resolution, which is an ideal nondestructive optical means for imaging and measuring the magnetism of nanometer thick and micrometer-size flakes. The geometry of the fiber-based zero-area loop Sagnac interferometer guarantees the rejection of reciprocal effects and senses only nonreciprocal effects with high precision, thus affording a static measurement of the absolute Kerr rotation angle without sample modulations. These merits make it ideal for detecting small signals from as-exfoliated 2D flakes without involving any device fabrication process, which in this work might perturb the magnetism. The detailed experimental setup can be referred to Fig. S6. It is important to note, in such a sensitive experiment, ~3.5 μm light spot is focused on nanometer thick and micrometer size flakes, any horizontal drifting and vertical defocusing would possibly induce large signal fluctuations. In order to avoid the drifting induced invariance, every Kerr rotation data point throughout this paper is a summary based on a scanned image including the target flake and the surrounding substrate, except the bulk measurement in Fig. 2j. Magnetic properties of bulk crystal have also been examined by superconducting quantum interference device (SQUID) in this work.

In order to observe ferromagnetism in few layers, we monitored the temperature dependent Kerr image of the 2L sample in Fig. 2a with the help of a small perpendicular field of 0.075 T to stabilize the magnetic moments. Strictly speaking, upon applying an external magnetic field, we may no longer have a well-defined ferromagnetic phase transition (at temperature $T_C$ with zero field), therefore we call it a transition between a ferromagnetic-like state and a paramagnetic-like state, separated by a transition temperature, $T_C^*$. Fig. 2b-e clearly shows the emergence of ferromagnetic order in bilayer $Cr_2Ge_2Te_6$ as temperature decreases: at 40 K, the Kerr intensity of the whole scanning area except for the region of thicker flakes ($\geq$ 3 layers) is hardly discernable; as temperature continues to drop, the recognition of the long strip is gradually enhanced; while approaching the liquid helium temperature, the bilayer long strip is clearly distinguishable from the surrounding bare substrate, in terms of the Kerr rotation angle intensity. The clear visibility of thicker ($\geq$ 3 layers) flakes at 40 K implies a higher $T_C^*$ for thicker crystals, which will be further discussed.

The non-zero transition temperature $T_C^*$ in the bilayer sample reflects a true 2D ferromagnetic order in our system. Firstly, the observed ferromagnetism is not due to substrate polarization because silicon dioxide is nonmagnetic and does not strongly bond to $Cr_2Ge_2Te_6$ layers. Secondly, the exposure-to-air time from the flake exfoliation to the specimen loading in a vacuum chamber ($10^{-6}$ torr) is strictly controlled to be under 15 minutes. The ambient effects, even if there were some, would generally reduce $T_C^*$. Therefore, the observation of finite $T_C^*$ is an unambiguous evidence of 2D ferromagnetism originating from $Cr_2Ge_2Te_6$ atomic layers.

A strong dimensionality effect is revealed by a thickness-dependent study under 0.075 T field. Fig. 2f-k display the prominent monotonic uplifting of $T_C^*$ with increased thickness, from a bilayer value ~30 K to a bulk limit ~68 K. The behavior of $T_C^*$ from 2D to 3D regimes is similar to the universal trend of many magnetic transition metal thin films,[25-27] although the interlayer bonding strength in vdW crystals is two-to-three orders of magnitude weaker than that of traditional metals. The strong thickness dependence of $T_C^*$ suggests the essential role of interlayer magnetic coupling in establishing the ferromagnetic order in $Cr_2Ge_2Te_6$ crystals. In other words, a ferromagnetic order in bulk vdW crystals does not guarantee a ferromagnetic order in 2D sheets, which highlights the necessity of a measurement directly on atomic layers as in the present work.

To gain a deeper level of understanding of the observed 2D ferromagnetic behavior in $Cr_2Ge_2Te_6$, we study a Heisenberg Hamiltonian with magnetic anisotropies, $H = \frac{1}{2}\sum_{i,j} J_{ij} \boldsymbol{S}_i \cdot \boldsymbol{S}_j +$

$\sum_i A(S_i^z)^2 - g\mu_B \sum_i B S_i^z$ ($\boldsymbol{S}_i$ : spin operator on site $i$, $J_{ij}$ : exchange interaction between site $i$ and $j$, $A$: the single-ion anisotropy, $g$: the Landé g-factor, $\mu_B$: Bohr magneton, and $B$: external magnetic field). To solve this spin Hamiltonian, we apply the renormalized spin wave theory (RSWT)[28] which includes magnon-magnon scatterings at the Hartree-Fock level self-consistently (SI). The input interaction parameters (three intralayer $J_\parallel$'s, three interlayer $J_\perp$'s and $A$) are calculated based on *ab initio* density functional method[29] at the local spin density approximation plus $U$ level (SI). We find small $U$ values ($U$ = 0.5 eV in this work) reproduce the magnetic ground state properties of bulk $Cr_2Ge_2Te_6$ (SI). The $J$'s directly mapped out from first-principles density functional calculations appear to be overestimated within RSWT, and are rescaled by a fixed and uniform factor of 0.72 that pins to the bulk $T_C$ (SI). We perform layer-dependent calculations including Zeeman effect from the 0.075 T external field (Fig. 2k) assuming $A = 0$ in few layers and $A = $ -0.05 meV in bulk (SI), and observe a strong dimensionality effect in line with experimental observations. At finite temperatures, spontaneous symmetry breaking (ferromagnetic order) takes place in 3D for the rotationally invariant isotropic Heisenberg system ($A = 0$ and $B = 0$), but is completely suppressed by the thermal fluctuations of the long-wavelength gapless Nambu-Goldstone modes in 2D. Nevertheless, magnetic anisotropy ($A \neq 0$ or $B \neq 0$) could establish ferromagnetic order in 2D at finite temperatures by breaking the continuous rotational symmetry in Hamiltonian, therefore giving rise to a nonzero excitation gap to the lowest energy mode of the acoustic magnon branch. Thermal energy at finite $T_C^*$ excites a large number of low-energy but finite-frequency magnon modes, flattening the expectation value of the collective spins. As the number of layer increases (from 2D to 3D), the increased thickness rapidly reduces the density of states per spin for the magnon modes near excitation gap (Fig. 1c-e), requiring a higher $T_C^*$ to create enough population of excitations to destroy the long-range magnetic order, thus exhibiting a strong dimensionality effect. This picture clearly distinguishes our observed 2D ferromagnetism in a real 2D material from the quasi-2D ferromagnetism embedded in bulk materials,[30,31] in which there is still weak out-of-plane ferromagnetic coupling and the magnon density of states is different (see Fig. 1).

The agreement between experiments and the theoretical calculations including Zeeman effect in the Hamiltonian reveals a strong role of the external field in the observed transition temperature. In order to pin down the effects of external field and intrinsic anisotropy, we did the experimental study of the hysteresis loop on the 6L sample at 4.7 K. The extreme softness (about 2% zero-field remanence of the maximum Kerr signal at 0.6 T) is observed, suggesting a very small intrinsic anisotropy (< 1 μeV, SI). Two zero-field Kerr images (Fig. 3b,c) after withdrawing the magnetic field from 0.6 T and -0.6 T show tiny yet definitive remanence with opposite signs (see line cuts in Fig. 3d,e). Kerr images under descending fields down to zero (Fig. 3f-k,b) show the flake is a single ferromagnetic domain through out. In ultrathin flakes, dipolar interaction is expected to be very small, and the equilibrium domain size which diverges exponentially with thickness would likely be very large.[32] The observation of about 2% single-domain remanence is an unambiguous evidence of the strong thermal fluctuations in 2D regime, in contrast to 3D ferromagnets where fractional remanence (if any) is usually caused by multi-domains. We also scanned the magnetic field on 2L and 3L flakes and did not see remanence within the detection limit.

The vanishing remanence in 3L at 4.7 K, together with the observation of $T_C^* = 41$ K under 0.075 T field, suggests a strong magnetic field control of transition temperatures in our 2D magnetic systems. A field that is usually deemed to be too small (e.g., < 0.5 T) to affect transition temperatures of 3D systems can drastically affect the behavior of 2D soft ferromagnetism by opening the spin wave excitation gap, as sketched in Fig. 1c-e. In order to further examine this scenario, we did temperature dependent Kerr rotation study on 2L, 3L, and 6L samples under two contrasting fields: 0.065 T and 0.3 T. From 0.065 T to 0.3 T, $T_C^*$ of 2L increases from 28 K to 44 K (57% enhancement), $T_C^*$ of 3L increases from 35 K to 49 K (40% enhancement), and $T_C^*$ of 6L increases from 48 K to 65 K (35% enhancement) approaching the bulk limit. In stark contrast, the $T_C^*$ in the bulk from SQUID measurements under 0.025 T to 0.03 T does not show clear change. Under the two fields (0.065 T and 0.3 T), the overall shift of magnetization-temperature curves in 2D layers (Fig. 4a-c) is obviously distinguished from the tail effect of a magnetic field on the critical region above $T_C^*$ in bulk (Fig. 4d). The possibility of superparamagnetism can be ruled out because of the integrity of the crystalline flake from high-quality single-phase crystals, evidenced by X-ray diffraction[23] and transmission electron microscopy measurements.[33]

The observed field effect on transition temperature can be explained well within the picture of spin wave excitations as illustrated in Fig. 1. In 2D, ferromagnetic transition temperature critically depends on the size of the spin wave excitation gap. In a 2D ferromagnet with negligible single-ion anisotropy, a magnetic field helps to increase the magnetic stiffness logarithmically, leading to a rapid increment of $T_C^*$, particularly in the range of small fields. However, in 3D, $T_C^*$ is predominantly determined by exchange interactions, and is insensitive to small single-ion anisotropies and small fields that are usually orders-of-magnitude smaller than $J$'s.

By means of RSWT method, we calculated the magnetic field dependence of $T_C^*$ in few-layers and bulk. The calculation results are quantitatively consistent with experimental values, as shown in Fig. 4. In all the 2L, 3L and 6L samples, a remarkable change of $T_C^*$ can be obtained in the range of the magnetic fields used in experiments (Fig. 4f-h), but not in the bulk (Fig. 4i). Considering the experimental challenge to determine the value of the tiny intrinsic anisotropy in a quantitative accuracy, we further investigated theoretically the effects of reasonably small but finite intrinsic anisotropies (see SI and Fig. S10), and find that the effect of field control on transition temperatures is still robust. Interestingly, for 6L under 0.3 T, the calculated $T_C^*$ approaches the bulk $T_C$, agreeing well with experiments. It thus provides an intriguing platform in which magnetic properties can be easily modulated between 2D and 3D regimes (at least in view of the transition temperature).

We report the first experimental observation of intrinsic ferromagnetism in 2D ferromagnetic vdW crystal $Cr_2Ge_2Te_6$, where a strong dimensionality effect arises from the low-energy excitations of magnons. Through the effective engineering of the magnetic anisotropy by small magnetic fields, we discover the unprecedented magnetic field control of transition temperatures in 2D soft ferromagnetic vdW crystal. Our experimental observations are confirmed by the rigorous renormalized spin wave theory analysis and calculations, which can provide generic understanding of the ferromagnetic behaviors of many 2D vdW soft ferromagnets such as $Cr_2Si_2Te_6$ and $CrI_3$. Our discovery of 2D soft ferromagnetic vdW crystal $Cr_2Ge_2Te_6$ provides a close-to-ideal 2D Heisenberg ferromagnet for exploring fundamental physics, and opens new possibilities for applications such as ultra-compact spintronics.

**Figure legends:**

**Figure 1. Schematics of spin wave excitations in two dimensions (2D) and three dimensions (3D). a,** Ferromagnetic spin wave (magnon) excitations in 2D, with intralayer exchange interaction $J_\parallel$, single ion anisotropy $A$, and magnetic field $B$. **b,** Ferromagnetic spin wave excitations in 3D, with extra interlayer exchange interaction $J_\perp$. **c-e,** Magnon density of states (DOS) around the low-energy magnon band edge of monolayer, multi-layer and bulk ferromagnetic materials, respectively. The low-energy excitations from the ferromagnetic ground state follow parabolic dispersions, and accordingly the DOS is a step function in 2D, and is $\sim \sqrt{E}$ in 3D, where $E$ is excitation energy. Therefore, more magnons are excited for a given thermal energy in 2D than in 3D due to the reduced dimensions. In 2D, the ferromagnetic transition temperature $T_C$ is primarily determined by the excitation gap, which results from the magnetic anisotropy, whereas in 3D, $T_C$ is primarily determined by exchange interactions. **f,** Crystal structure (side view and top view) of $Cr_2Ge_2Te_6$. Bulk $Cr_2Ge_2Te_6$ has layered structure, and the layers are coupled through vdW interaction with a 3.4 Å interlayer spacing.

**Figure 2. Observation of ferromagnetism in bilayer (2L) $Cr_2Ge_2Te_6$ and temperature dependent Kerr rotation of few-layer and bulk $Cr_2Ge_2Te_6$. a,** The optical image of exfoliated $Cr_2Ge_2Te_6$ atomic layers on 260 nm $SiO_2$/Si, consisting of a 31 μm long bilayer strip attached to a trilayer flake with an even thicker end. The bar is 10 μm. **b-e,** The emergence of a Kerr rotation signal for the bilayer flake upon the application of a small magnetic field of 0.075 T, as the temperature decreases from 40 K to 4.7 K. As the temperature drops, the Kerr rotation signal from the bilayer flake increases, making the bilayer region in sharp contrast with the nearby bare $SiO_2$ substrate. The clear visibility of thicker flakes (≥ 3L) indicates a higher $T_C$. The average background signal is subtracted (Figure S6) and the signals of higher intensities from the thicker flakes (≥ 3L) are truncated at 30 μrad. **f-j,** The temperature dependent Kerr rotation intensities of 2L, 3L, 4L, 5L and bulk samples under 0.075 T field. Error bar represents the standard deviation of sample signals. **k,** Transition temperature $T_C^*$ (defined in the presence of external magnetic field) between a ferromagnetic-like state and a paramagnetic-like state in sample with different thickness, obtained from Kerr measurements and renormalized spin wave theory calculations. A strong dimensionality effect is seen, owing to the rapid reduction of onset magnon DOS as the thickness increases. The transition temperature $T_C^*$ is determined experimentally in the range

approximating the paramagnetic tail of the effective Kerr signal, and theoretically by the vanishing net magnetization. The statistics yield a finite uncertainty in the deterministic experimental values.

**Figure 3. Ferromagnetic hysteresis loop with single domain remanence in a six-layer (6L) Cr$_2$Ge$_2$Te$_6$. a,** Hysteresis loop of a six-layer flake at 4.7 K showing saturating trend at 0.6 T and tiny non-vanishing remanence at zero field. The solid red (blue) arrow represents the descending (ascending) field from the positive (negative) maximum. Loop starts from positive maximum 0.6 T. Inset is an optical image of the flake and the bar is 10 μm. **b, c,** Scanned Kerr rotation images of the flake at zero field after withdrawing the field from 0.6 T and -0.6 T, respectively. **d, e,** Line scanning crossing the flake with long acquisition time (each physical spot with 100 data acquisitions). The approximate line positions are indicated by black dashed lines in **b** and **c**, respectively, but extending further out of the window **b** and **c**. **b-e** show a clear signature of small but definitive remanent Kerr rotation angles (about 2% of the "saturated" Kerr rotation at 0.6 T) with opposite signs on ascending and descending branches. **f-k,** Kerr images of the highlighted area in **a** (within dashed rectangle), at different descending fields, showing the persistence of a magnetic single domain through out. The tiny remanence in a single domain is strong evidence that the effect of thermal fluctuations rather than the formation of multi-domains (usually in the bulk) is the reason of the reduced magnetization in our few-layer samples.

**Figure 4. Magnetic field control of transition temperature of few-layer Cr$_2$Ge$_2$Te$_6$. a-c,** Normalized Kerr rotation angle as a function of temperature, under two different magnetic fields of 0.065 T and 0.3 T, for 2L, 3L and 6L, respectively. The 0.3 T field shifts the curve drastically, with respect to curve under 0.065 T field, indicating a strong renormalization of the transition temperature $T_C^*$ in atomic layered Cr$_2$Ge$_2$Te$_6$. Experimental error bars are smaller than the plotted dot size if not shown. **d,** Temperature dependent magnetization curves of bulk crystals measured by SQUID under 0.025 T and 0.3 T. Compared with the curve under 0.025 T, the 0.3 T field only introduces a slightly distorted tail above $T_C^*$. The different behavior below $T_C^*$ is a possible result from domains: under 0.025 T field, multi-domains are likely formed (suggested by reduced absolute magnetization); but under 0.3 T, single-domain has been approached (see Fig. S1). **e-i,** Experimental and theoretical field dependence of $T_C^*$ in samples of various thickness. In **f** and **g**, experimental results at 0.075 T (Fig. 2) are plotted as well. In the 2D limit, if the single ion

anisotropy is negligibly small, the transition temperature will be very low, and can be easily tuned with a small magnetic field (e.g., B < 0.5 T). In the bulk limit, due to the 3D nature, such tuning ability does not hold. Due to the large signal and the continuous temperature sweeping, all $T_C^*$ values of the bulk crystal measured by SQUID are determined at the steepest slop of the magnetization-temperature characteristic curve.

**Acknowledgements:**

We thank Dr. J.-G. Zheng for helping with AFM measurement at UC Irvine. This work was primarily supported by the Director, Office of Science, Office of Basic Energy Sciences, Materials Sciences and Engineering Division of the US Department of Energy under Contract No. DE-AC02-05-CH11231 (van der Waals heterostructures program, KCWF16). The bulk crystal growth at Princeton University was supported by the NSF MRSEC program, grant NSF DMR-1420541, and part of measurement at UC Irvine was supported by NSF DMR-1350122.

**Author Contributions:**

C.G. and X.Z. conceived and initiated this research and designed the experiments; C.G. performed the Sagnac MOKE measurements with assistance from L.L., A.S. and Y.X., under the guidance of J.X.; Z.L. and T.C. performed theoretical calculations under the guidance of S.G.L.; H.J. synthesized $Cr_2Ge_2Te_6$ bulk crystals under the guidance of R.J.C.; C.G. prepared and characterized few-layer samples with assistance from Y.X., W.B. and C.W.; C.G., Z.L., T.C., Y.W., Z.Q.Q., S.G.L., J.X. and X.Z. analyzed the data and wrote the paper.

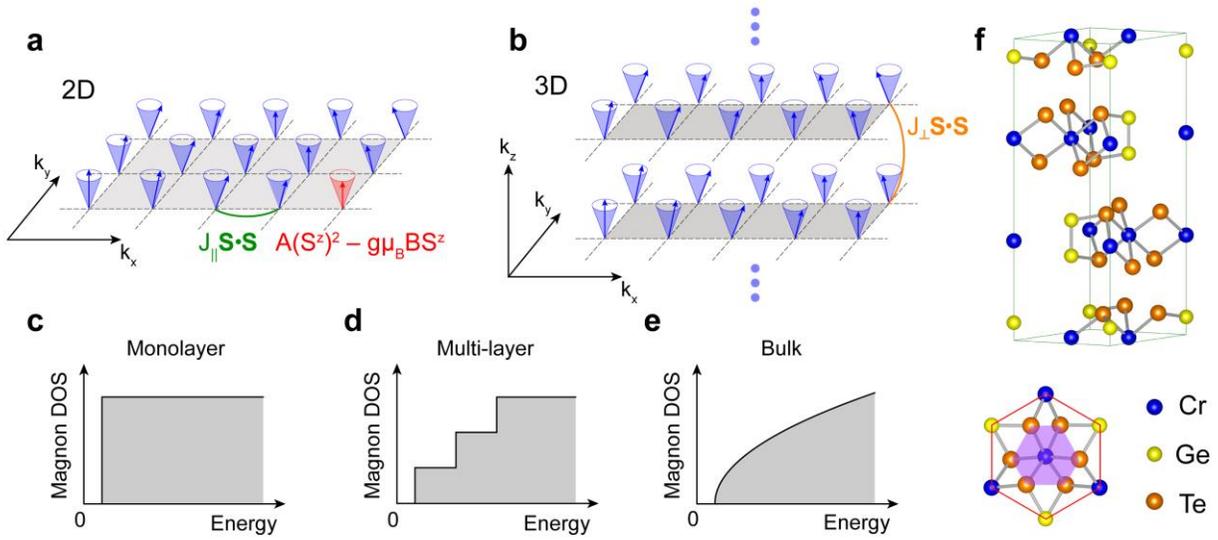

**Figure 1**

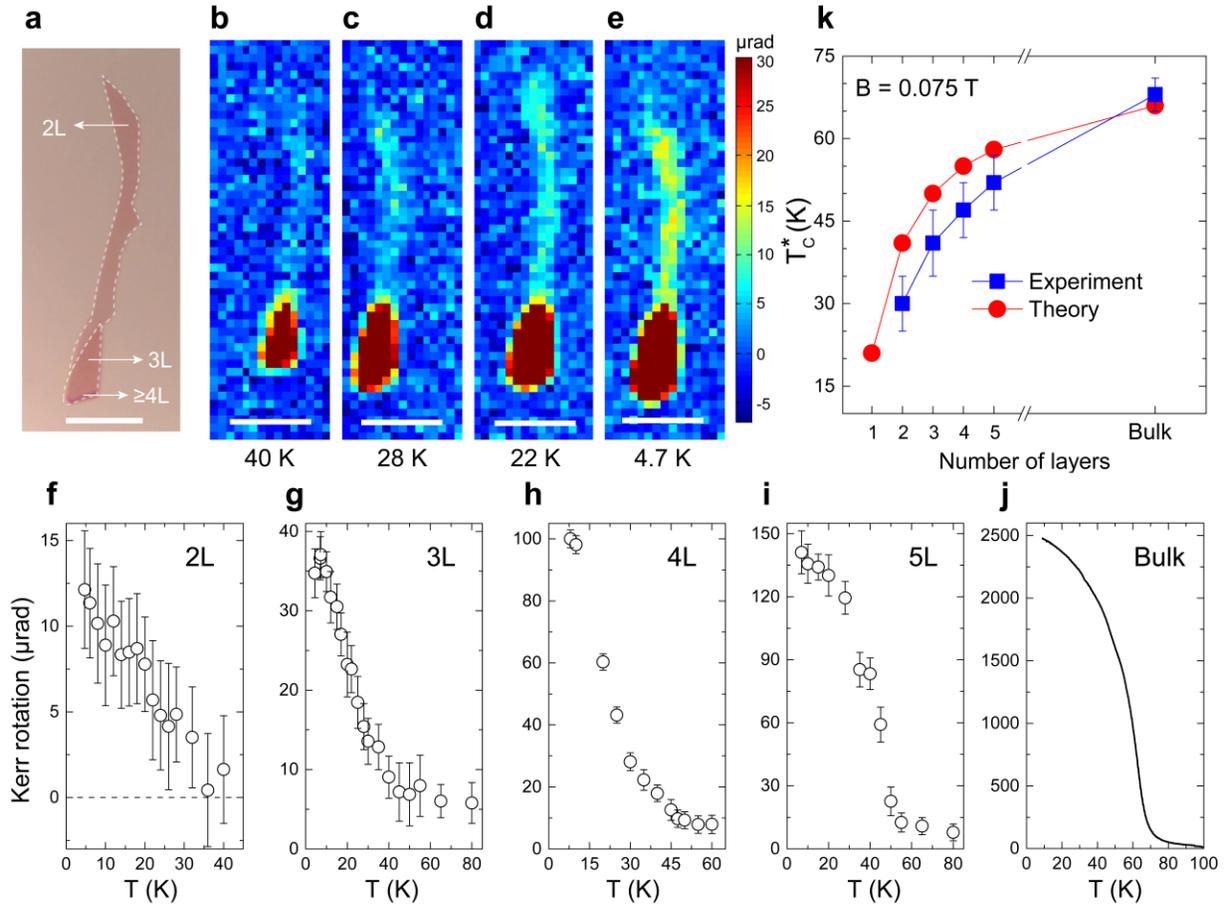

**Figure 2**

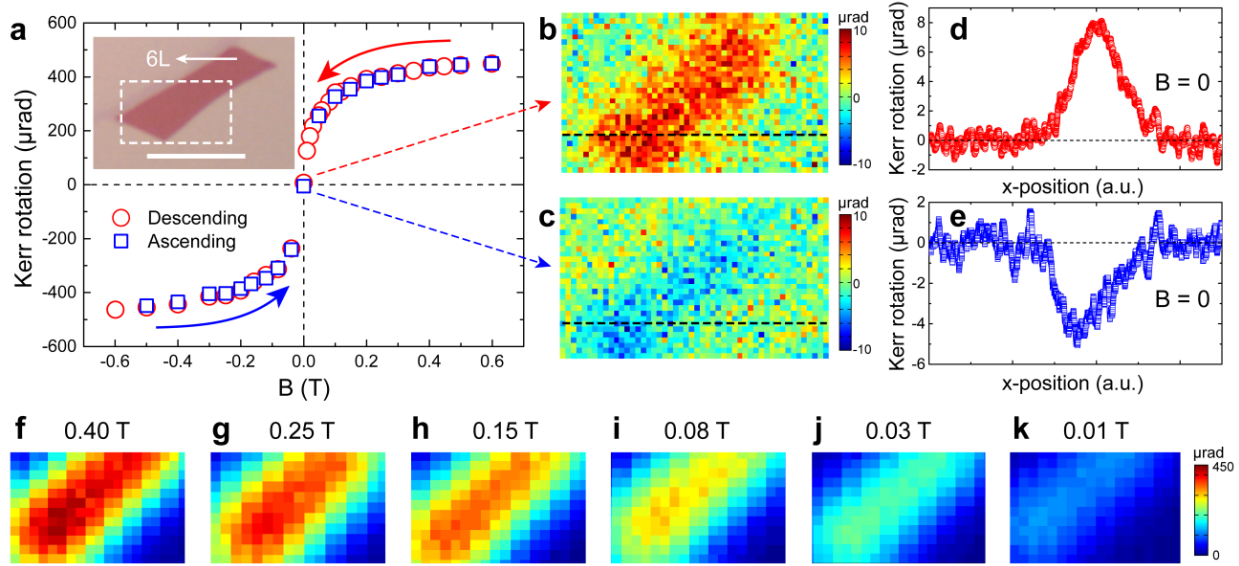

**Figure 3**

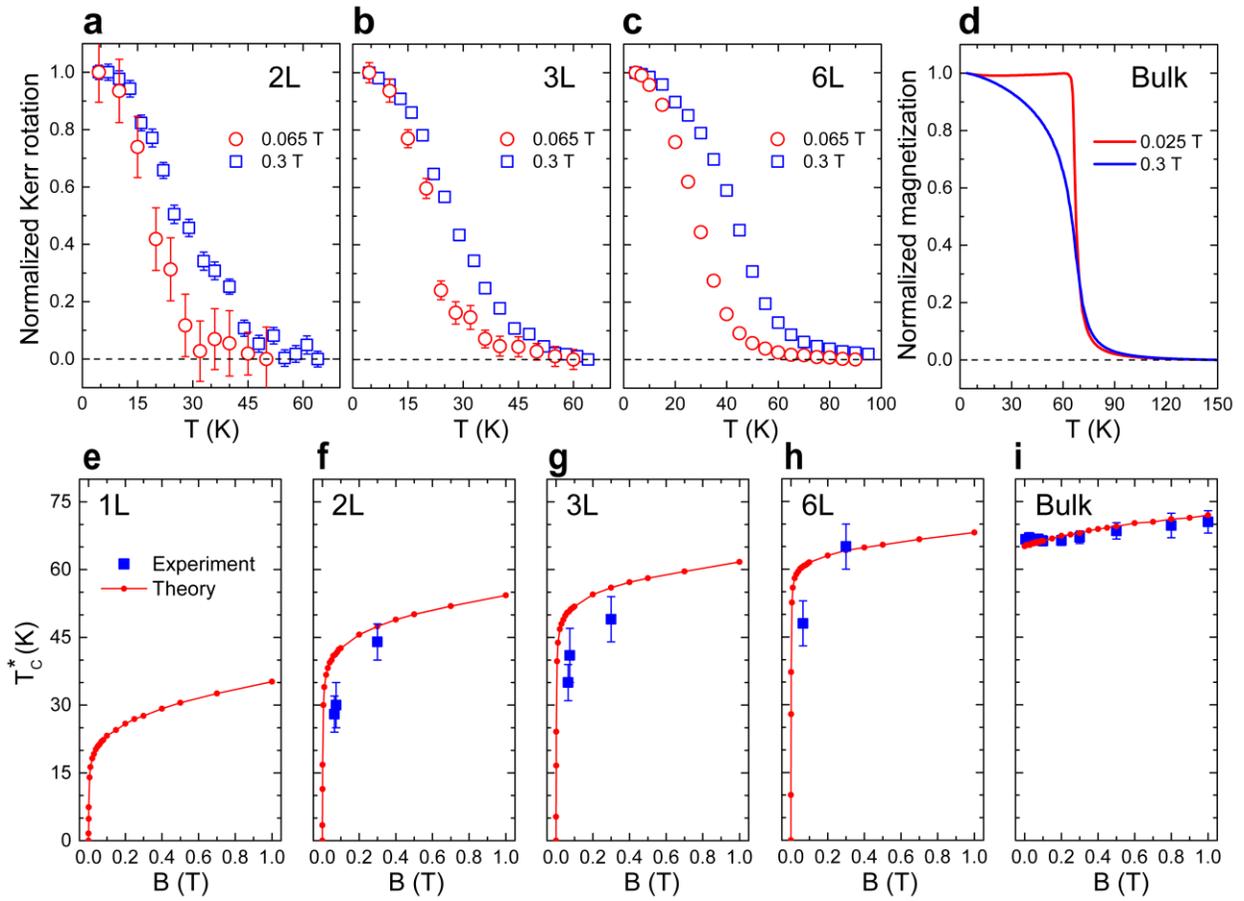

**Figure 4**